\documentclass[conference]{IEEEtran}
\IEEEoverridecommandlockouts
\usepackage{cite}
\usepackage{amsmath,amssymb,amsfonts}
\usepackage{booktabs}
\usepackage{algorithmic}
\usepackage{graphicx}
\usepackage{textcomp}
\usepackage{xcolor}
\usepackage{multirow} 
\usepackage{tikz}
\usepackage{pgfplots}
\usepackage{threeparttable}

\def\BibTeX{{\rm B\kern-.05em{\sc i\kern-.025em b}\kern-.08em
    T\kern-.1667em\lower.7ex\hbox{E}\kern-.125emX}}
\begin{document}

\title{Contrastive Learning for Aspect Representation towards Explainable Recommendation}

\author{
Emrul Hasan\\
 Department of Computer Science\\
 Toronto Metropolitan University\\
 Toronto, Canada\\
  \texttt{e1hasan@torontomu.ca}
  \and
Chen Ding\\
 Department of Computer Science\\
 Toronto Metropolitan University\\
 Toronto, Canada\\
  \texttt{cding@torontomu.ca}
}

\maketitle

\begin{abstract}

In this work, we propose a novel recommendation model, CLARER (\textbf{C}ontrastive \textbf{L}earning for \textbf{A}spect \textbf{R}epresentation
towards \textbf{E}xplainable \textbf{R}ecommendation) that integrates aspect features learned from textual reviews with rating information to improve the accuracy and explainability of recommendations. Our proposed framework learns user and item representations by combining rating-based features and aspect-based features from reviews. Specifically, rating-based features are learned through a multi-layer perceptron (MLP) model, while aspect-specific review representations are learned using a transformer encoder to capture the semantic information and contrastive learning to better distinguish user preferences. To provide explanations, we train a transformer decoder, using the final representations of users and items from both rating and aspect-based features as context. Experimental results in three benchmark data sets demonstrate that our model achieves superior performance compared to baseline methods in both recommendation (accuracy) and explanation generation. 

\end{abstract}
\section{Introduction}
While recommendation systems are essential for delivering personalized suggestions in today's digital landscape, traditional approaches relying on latent features from past interactions often lack interpretability \cite{he2017neural,sarwar2001item,zhang2019deep,zheng2017joint,xu2021multi}. To enhance the credibility and user acceptance of the recommendation results, it is essential to provide clear justifications or explanations for the recommendations \cite{zhang2020explainable,tintarev2015explaining}. Justifications can be expressed in several different forms, e.g., using a predefined template \cite{zhang2014explicit,li2021caesar}, highlighting the important parts \cite{chen2019personalized}, generating short descriptions \cite{chen2021generate,li2020generate}. Given the significance of explainable recommendations, this crucial area of research has received considerable attention in recent years \cite{liao2025aspect, wang2018explainable, ma2019jointly}.

One of the main challenges in explainable recommendation systems is to identify and incorporate more contextual information and fine-grained user preferences and opinions into personalization and explanation generation. Many existing approaches rely on user and item ID-based feature representation and use them as contextual information in the explanation generation process \cite{li2017neural,dong2017learning,li2020generate}. However, ID-based features exist in a semantic space different from that of natural language, leading to poor alignment between two different semantic spaces and limited personalization in explanation generation. Furthermore, due to the sparsity of historical rating data, ID-based features often fail to capture accurate representations of users and items \cite{natarajan2020resolving}. As a result, incorporating only ID-based features into the recommendation model performs suboptimally both in rating prediction and in explanation generation.

To address these issues, previous research has attempted to map user and item IDs onto textual representations and integrate them into the decoding process for explanation generation \cite{li2017neural}. Although this approach introduces a degree of personalization, it also increases the model complexity and suffers from inconsistencies in mapping ID-based features to natural language explanations. Consequently, there remains a need for a more effective approach that could directly integrate user preferences into the recommendation generation process while ensuring interpretability and coherence.

Another major challenge is the feature extraction from user reviews. Capturing fine-grained user preferences from historical interactions is crucial to improving recommendation performance. User-generated reviews provide the valuable information, but extracting meaningful features from textual data remains difficult due to its unstructured and noisy nature. Some recent approaches have employed external aspect extraction tools to pre-process textual data before incorporating them into the explanation generation model \cite{le2020synthesizing}. Although these methods have shown promise, they rely heavily on external tools, making them less adaptable across different domains. Thus, developing a self-contained feature extraction mechanism that effectively captures aspect-based user preferences remains an open challenge.

To overcome limitations of existing explainable recommendation methods, we propose CLARER (\textbf{C}ontrastive \textbf{L}earning for \textbf{A}spect \textbf{R}epresentation towards \textbf{E}xplainable \textbf{R}ecommendation), a novel framework that jointly integrates ID-based features with aspect-aware representations learned from user reviews. The key contribution of our work can be summarized as follows.

    1) We propose CLARER, a novel explainable recommendation framework that seamlessly integrates aspect-based representations with user-item latent factors, improving both rating prediction and explanation generation. To the best of our knowledge, CLARER is the first to combine ID-based features with aspect representations for both rating prediction and explanation generation (i.e., the decoder is initialized using the combined feature).
    
   2) We introduce a novel aspect representation learning technique that uses predefined domain-specific aspects and contrastive learning. We are the first to apply contrastive learning for aspect representation in explainable recommendation systems. 
   
   3) We run experiments on three real-world datasets and the results show that our approach outperforms state-of-the-art baselines in both accuracy and explainability, with aspect-aware representations driving the performance gains.


\section{Related work}
This section presents two key areas relevant to our work: aspect representation and explainable recommendation systems.

\subsection{Aspect Representation}
Deep learning, particularly representation learning, is widely used to automatically capture meaningful features from reviews. For example, CNN-based models \cite{seo2017interpretable, zheng2017joint} extract features from the review text, while attention mechanisms prioritize important reviews and integrate rating-based and review-based features. Notable models include NARRE \cite{chen2018neural}, which evaluates the usefulness of the review, and mutual attention methods \cite{liu2019daml} that capture dynamic interactions. Hierarchical attention networks \cite{dong2020asymmetrical} distinguish user and item structures, while gating mechanisms \cite{xia2019leveraging} adaptively balance ratings and reviews. Further advances, such as MSAR \cite{peng2020mutual} and TARMF \cite{lu2018coevolutionary}, improve feature fusion using gated layers and recurrent neural networks. 

Although various topic modeling and attention-based models are applied to general review-based or aspect-based recommendation systems, aspect-specific representations for explainable recommendations have yet to be explored fully. \cite{bao2014topicmf} utilizes non-negative matrix factorization \cite{lee1999learning} to extract latent topics from review texts while applying matrix factorization to model rating data. Topic-based methods, such as Latent Dirichlet Allocation (LDA), are used to model aspects from reviews for recommendation tasks \cite{mcauley2013hidden, diao2014jointly} and extended with aspect-aware models by \cite{cheng2018aspect}. These topic-based models integrate learned topics into rating-based frameworks, such as matrix factorization, enhancing the interpretability of recommendation systems. However, these methods primarily rely on bag-of-words representations, overlooking comprehensive semantic information from reviews, which limits their ability to fully capture meaningful insights.
 
Very recently, \cite{liao2025aspect} incorporates aspect representation as the context for explanation generation. They extract relevant aspects from the review, which are then encoded using the NLP technique. We argue that the dependency on an external tool might restrict the ability of the model to capture all the aspects that users discuss in the review, leading to suboptimal representation learning. To this end, we propose using contrastive learning for predefined aspect representation learning. This technique contrasts the individual aspect with the entire representation of the review. Therefore, the aspect representations are rich, leading to performance improvement. To the best of our knowledge, we are the first group to use this technique in building explainable recommendation systems.

\vspace{-0.5em}
\subsection{Explainable Recommendation Systems}
 One of the main challenges in generating personalized explanations is incorporating contextual information into explanation generation \cite{chen2018neural,chen2021generate,chen2017explaining,chen2019personalized}. The traditional approach relies on the user and item ID-based feature representation. For instance, \cite{li2017neural} leverages historical rating data to learn user and item representations for personalization. In addition, this model is trained to generate reviews by integrating both user-item representations and review features as contextual information for explanation generation. Similarly, \cite{li2020generate} utilizes historical user-item rating data to model user-item interactions, using them to initialize the decoder network for template-controlled explanation generation. ID-based feature representation models face challenges in handling sparse user-item interactions.

Meanwhile, \cite{dong2017learning} employs historical interaction data and an attention based sequence-to-sequence model for attribute-conditioned review generation. Expanding on the role of IDs in explanation generation, \cite{li2023personalized} incorporates IDs as prompts to enhance explanation generation using large language models (LLMs). Although LLM-based models demonstrate promising potential, they struggle to effectively capture user-item collaborative preferences, particularly complex and nonlinear relationships between users and items. Recently, some works have focused on using textual reviews as contextual information to initialize the decoder model for explanation generation. For instance, \cite{wang2018explainable} integrates the opinionated review text to enhance the recommendation and explanation. \cite{liao2025aspect} integrates multi-modal information, e.g., ID-based features, aspect features, retrieval sentences, and interpretation text, improving the quality of the explanation. On the other hand, \cite{tang2024gavamoe} uses a variational autoencoder with GMM for rating reconstruction and a multi-gating mixture of experts for explanation generation, increasing personalized recommendation and effectively handling the data sparsity issue. Despite their effectiveness, LLM-based approaches convey out-of-domain information inherently since they are trained on the general internet data.  In addition, they ignore integrating the strength of both rating and aspect-based features for rating prediction and explanation generation. To this end, we propose to integrate aspect-specific features for both rating prediction and explanation generation.

\section{Methods}
In this section, we present our proposed framework, CLARER, as illustrated in Fig. \ref{fig: architecture}, and provide a detailed description of its three main components.

\subsection{Priliminaries}

Let us consider a user $u$, an item $v$, a review $x$, and an explanation $e$ (during training only), the objective is to predict the overall rating $r_{uv}$, which reflects the user's preference for the item, along with an explanation $e\prime$ that justifies why the item is recommended.

\subsection{User and Item Representation}

We employ a simple yet effective method for user and item ID-based representation using a multilayer perceptron (MLP). While MLP-based ID representations have certain limitations, they are favored in our approach due to their simplicity, computational efficiency, and ability to model complex non-linear interactions \cite{he2017neural}. We begin by randomly initializing the embeddings for users and items. These embeddings are concatenated and propagated through a series of hidden layers. The final output of the hidden layer is combined with the aspect-level representation derived from user-generated reviews, as discussed in section III (C). The combined feature is fed into a linear layer to generate a scaler value, e.g., a rating score, which is optimized by minimizing the difference from the ground-truth rating.

Formally, consider \( p_u \) and \( q_v \) to be the latent representations of the user \( u \) and the item \( v \) respectively.  To learn the representation, the latent vectors are concatenated as follows.

\begin{equation}
    z = \text{Concatenate}(p_u, q_v)
\end{equation}

This concatenated vector is the input of the MLP model, e.g., \( h_0 = z \). The hidden layer $h_{l}$ and the final layer are defined as follows.
\begin{equation}
    h_l = \text{ReLU}(W_l h_{l-1} + b_l)
\end{equation}
\begin{equation}
    h_L = \text{ReLU}(W_L h_{L-1} + b_L)
\end{equation}

Here, $W_{l}$ and $W_{L}$ denote the weight matrices for the hidden layer and the final layer respectively, while $b_{l}$ and $b_L$ represent the bias terms for the $l$-th and $L$-th layer, respectively. $ReLU$ is the activation function.

\begin{figure}[ht]
\centering
\includegraphics[width=\columnwidth]{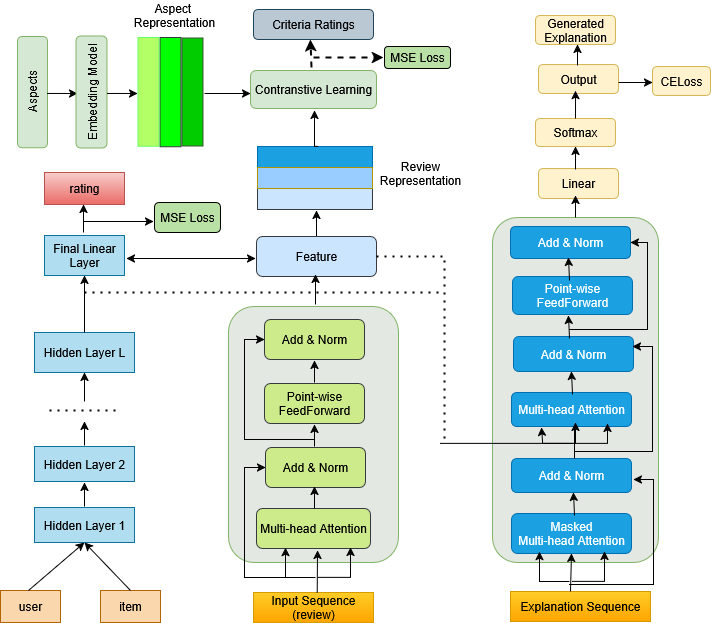}
\caption{The detailed architecture of the proposed model.}
\label{fig: architecture}
\end{figure}

\subsection{Aspect Representation}

Aspect representation learning involves learning the rich representation of various aspects discussed in the review. To this end, we utilize the contrastive learning method \cite{le2020contrastive} to learn representations of reviews at the aspect level and combine them with ID-based features to predict ratings and generate explanations. In our approach, we consider domain-dependent aspects. The details of the predefined aspects for each domain are shown in Table \ref{tab:aspects_domains}. The aspects in the TripAdvisor dataset are predefined by the TripAdvisor platform, whereas for Amazon and Yelp reviews, we select and define aspects through data analysis, e.g., topic analysis.

\begin{table}[h]
    \caption{Aspect categories across domains}
    \centering
    \begin{tabular}{|p{6cm}|p{1.6cm}|}
        \hline
        \textbf{Aspects} & \textbf{Domain} \\
        \hline
        Plot, Storyline, Cultural Relevance, Emotion, Humor, Acting, Direction, Visuals, Sound, Genre, Characters, Entertainment & Amazon (MT) \\
        \hline
        Service, Food, Ambience, Cleanliness, Price, Variety, Location, Wait Time, Noise & Yelp \\
        \hline
        Service, Location, Cleanliness, Sleep, Room Size, Check-in, Dining, Value, Amenities, Safety & TripAdvisor \\
        \hline
    \end{tabular}
    \label{tab:aspects_domains}
\end{table}

The process starts with obtaining embeddings for all predefined aspects using a pre-trained LLM embedding model; here we use a sentence transformer \cite{reimers2019sentence}. Next, we train a transformer encoder to generate the review representation. Finally, the review features are contrasted with the aspect embeddings to learn the final review-based aspect-level representation.

For a given set of aspects $a_{1}, a_{2},...,a_{K}$ where $K$ is the number of aspects, the representation of an aspect $a_{k}$ can be expressed as
\begin{equation}
    a^{e}_{k}=Embedding(a_{k})
\end{equation}

To effectively extract meaningful features from textual reviews, we train a transformer encoder model \cite{vaswani2017attention}. The encoder processes user-generated reviews to learn rich, context-aware representations. The model uses self-attention mechanisms to capture long-range dependencies and semantic relationships within the text, enabling a fine-grained understanding of the review content. Once the review features are extracted, they are contrasted with aspect embeddings and trained using contrastive loss. For each aspect, the review feature is treated as an augmented version that better aligns with the aspect, while negative samples are generated from the aspect's embedding using a random generator. We consider the review as the positive counterpart of the aspect, assuming that the aspect is inherently reflected in the review, allowing the model to learn this relationship effectively.

Given an input review sequence of tokens \( \mathbf{X} = (x_1, x_2, \dots, x_T) \), the encoder processes the input through multiple layers consisting of self-attention and feed-forward networks. Each input token is first mapped to an embedding space and then combined with positional encodings. The combined feature $E$ is represented as follows:

\begin{equation*}
    \mathbf{E} = \text{Embedding}(\mathbf{X}) + \text{PositionalEncoding}(\mathbf{X})
\end{equation*}
\begin{equation*}
    \mathbf{Q} = \mathbf{E} W^Q, \quad \mathbf{K} = \mathbf{E} W^K, \quad \mathbf{V} = \mathbf{E} W^V
\end{equation*}
\begin{equation*}
    \text{Attention}(\mathbf{Q}, \mathbf{K}, \mathbf{V}) = \text{softmax} \left( \frac{\mathbf{Q} \mathbf{K}^T}{\sqrt{d_k}} \right) \mathbf{V}
\end{equation*}
\begin{equation*}
    \text{MultiHead}(\mathbf{Q}, \mathbf{K}, \mathbf{V}) = \text{Concat}(\text{head}_1, \dots, \text{head}_h) W^O
\end{equation*}
\begin{equation*}
    \mathbf{Z} = \text{LayerNorm}(E  + \text{MultiHead}(\mathbf{Q}, \mathbf{K}, \mathbf{V}))
\end{equation*}
\begin{equation*}
    \text{FFN}(\mathbf{Z}) = \text{ReLU}(\mathbf{Z} W_1 + b_1) W_2 + b_2
\end{equation*}
\begin{equation*}
    \mathbf{Z}_{o} = \text{LayerNorm}(\mathbf{Z} + \text{FFN}(\mathbf{Z}))
\end{equation*}

$\mathbf{Z}_{o}$ is the contextual representation of the review. We train this encoder module with contrastive loss to obtain the aspect-level representation. 

Let $\mathbf{Z}_{o} \in \mathbb{R}^d$ be the review embedding,  
$a^{e}_{k} \in \mathbb{R}^d$ be the positive samples (e.g. equation 4),  
$a^{e}_j$ be a randomly sampled negative aspect embedding from the current batch and  
$\tau$ be the temperature parameter. The contrastive loss $\mathcal{L}_{i,j}$ can be defined as

\begin{equation}
\mathcal{L}_{i,j} = - \log \frac{1}{K}\sum_{i=1}^{K} \frac{\exp( \text{sim}(a^{e}_k, Z_0) / \tau )}{\sum_{j=1}^{2N} \mathbb{1}_{[j \neq i]} \exp( \text{sim}(a^{e}_k, a^{e}_j) / \tau )}
\end{equation}
where $K$ and $N$ are the total number of aspects and the number of negative samples respectively. Total loss is obtained by averaging over all the aspects.

\subsection{Integration of ID-based Features with Aspect Representation}
First, ID-based user and item representations are learned using MLP, as discussed in Section III (B). Then, aspect-based features are incorporated using a gating function, defined as:
\begin{equation}
\chi = \sigma(W\cdot Z_o + b)*Z_o + (1 - \sigma(W\cdot Z_o + b)) * h_{L})
\end{equation}
where $Z_o$ represents the aspect-based features and $h_{L}$ denotes the ID-based representation of users and items. The sigmoid activation function $\sigma(\cdot)$ acts as the gate function that controls the contribution of two different representations. The final user-item representation, $\chi$, effectively encodes both the interaction dynamics and the aspect-specific information.

\subsection{Rating Prediction}
The recommendation can be defined as a rating prediction problem where the goal is to predict a rating score $r_{uv}$ based on the representation of the user $u$ and the item $v$. Once we obtain the combined feature representation from the ratings and reviews, we apply a simple regression layer to predict the ratings as follows.
\begin{equation}
    \hat{r}_{u,i} = \sigma(\mathbf{W}^r \chi + \mathbf{b}^r) + b^r
\end{equation}
where $\mathbf{W}^r \in \mathbb{R}^{d \times d}$, $\mathbf{b}^r \in \mathbb{R}^d$, and $b^r \in \mathbb{R}$ are weight parameters, and $\sigma(\cdot)$ is the sigmoid function. The task is trained with Means Squared Error loss as: 
\begin{equation}
    \mathcal{L}_{r} = \frac{1}{|\mathcal{T}|} \sum_{(u,i) \in \mathcal{T}} (r_{u,i} - \hat{r}_{u,i})^2
\end{equation}

where $r_{u,i}$ is the ground truth rating and $\mathcal{T}$ denotes the total number of samples. 

\subsection{Explanation Generation}

Inspired by the performance of the autoregressive language model \cite{vaswani2017attention}, we use a transformer-based decoder model for explanation generation. Given an input sequence of tokens \( \mathbf{Y} = (y_1, y_2, \dots, y_T) \), the decoder generates the output sequence autoregressively by attending to both the previous user-item representation generated from review and ID and the output of the encoder. The decoder first applies masked self-attention to prevent attending to future tokens. Given the user-item representation $\chi$, the key and value are updated as: 

\begin{equation}
    P(y_t | y_{<t}, \chi) = \text{softmax}(\mathbf{Z^\prime_o} W^O + b)
\end{equation}

where \( W^O \) and \( b \) are learned parameters. $\mathbf{Z^\prime_o}$ is the decoder output which follows a similar operation as discussed for the encoder module in Section III(C). This process is repeated for each token until the last token is generated. The model is trained using the cross-entropy loss and the total explanation loss $\mathcal{L}_e$ is defined as follows: 

\begin{equation}
\mathcal{L}_e 
= \frac{1}{\lvert \tau \rvert} 
  \sum_{(u,i)\in \tau} 
    \frac{1}{\lvert S_{u,i} \rvert}
    \sum_{t=1}^{\lvert S_{u,i} \rvert} 
      -\log \, p(y_t)
\end{equation}
where $\tau$, and $S_{u,i}$ are the training set and the ground-truth explanation for user-item pair $(u,i)$.

\subsection{Multi-task Learning}
The core tasks of our framework are rating prediction and explanation generation. The individual training task ignores the influence of personalization in the explanation generation \cite{ma2019jointly}. Thus, we propose a multi-task learning objective function that trains all the tasks simultaneously. 

\begin{equation}
    O = \lambda_{1}
 \mathcal{L}_{r} +\lambda_{2}
 \mathcal{L}_{i,j} + \lambda_{3}
\mathcal{L}_e  +\lambda_n \|\Theta\|^2
\end{equation}
where $\Theta$ is the set of model parameters, and $\lambda_1$, $\lambda_2$, and $\lambda_3$ are regularization weights for user-item representation, aspect learning, and explanation generation, respectively. The regularization strength is controlled by the hyperparameter $\lambda_n$.

\section{Experiments}
\subsection{Dataset}
Three real-world datasets from different domains are used to evaluate the proposed model. Hotel ratings and reviews are collected from TripAdvisor\footnote{https://www.tripadvisor.com/}. In this work, we use one from the Hong Kong (HK) region \cite{li2020generate}. For the restaurant domain, the Yelp Challenge 2019 dataset\footnote{https://business.yelp.com/data/resources/open-dataset/} is used. Movie review dataset is sourced from Amazon’s 5-core Movies \& TV dataset (MT)\footnote{http://jmcauley.ucsd.edu/data/amazon}. We filter each dataset by removing users and items with fewer than 20 interactions to ensure sufficient information. Each record includes a user ID, item ID, rating (1–5), explanation, item feature, and textual review. For explanation generation, one sentence per review containing a feature is selected as the ground-truth. Key statistics are shown in Table~\ref{tab:dataset_stats}.

\begin{table}[h]
    \centering
    \begin{threeparttable}
    \caption{Statistics of datasets.}
    \begin{tabular}{lccc}
        \toprule
        & \textbf{TA-HK} & \textbf{YELP19} & \textbf{AZ-MT} \\
        \midrule
        \# of users & 9,765 & 27,147 & 7,506 \\
        \# of items & 6,280 & 20,266 & 7,360 \\
        \# of reviews & 320,023 & 1,293,247 & 441,783 \\
        \# of features & 5,069 & 7,340 & 5,399 \\
        Avg. \# of reviews / user & 32.77 & 47.64 & 58.86 \\
        Avg. \# of reviews / item & 50.96 & 63.81 & 60.02 \\
        Avg. \# of words / explanation & 13.01 & 12.32 & 14.14 \\
        \bottomrule
    \end{tabular}
    \begin{tablenotes}
        \footnotesize
        \item[*] TA and AZ denote \textbf{TripAdvisor} and \textbf{Amazon}, respectively.
    \end{tablenotes}
    \label{tab:dataset_stats}
    \end{threeparttable}
\end{table}

\begin{table}[ht]
    \centering
    \tiny
    \renewcommand{\arraystretch}{1.2}
    \setlength{\tabcolsep}{8pt}
    \begin{threeparttable}
    \caption{Performance on rating prediction using MAE and RMSE}
    \begin{tabular}{lcc|cc|cc}
        \toprule
        & \multicolumn{2}{c|}{\textbf{Yelp}} & \multicolumn{2}{c|}{\textbf{Amazon}} & \multicolumn{2}{c}{\textbf{TripAdvisor}} \\
        \cmidrule(lr){2-3} \cmidrule(lr){4-5} \cmidrule(lr){6-7}
        & RMSE $\downarrow$ & MAE $\downarrow$ & RMSE $\downarrow$ & MAE $\downarrow$ & RMSE $\downarrow$ & MAE $\downarrow$ \\
        \midrule
        PMF    & 1.097 & 0.883 & 1.235 & 0.913 & 0.870 & 0.704 \\ 
        NARRE  & 1.028 & 0.791 & 1.176 & 0.865 & 0.796 & 0.612 \\ 
        DAML   & 1.014 & 0.784 & 1.173 & 0.858 & 0.793 & 0.617 \\ 
        RGCL   & 1.008 & 0.784 & 1.160 & 0.872 & 0.791 & 0.611 \\ 
        NRT    & 1.016 & 0.796 & 1.188 & 0.853 & 0.797 & 0.611 \\ 
        CAML   & 1.036 & 0.798 & 1.191 & 0.888 & 0.818 & 0.622 \\ 
        PETER  & 1.017 & 0.793 & 1.181 & 0.863 & 0.814 & 0.635 \\ 
        SERMON & 1.002 & 0.784 & 1.159 & 0.837 & 0.791 & 0.599 \\ 
        CLARER & \textbf{0.716} & \textbf{0.509} & \textbf{0.999} & \textbf{0.700} & \textbf{0.642} & \textbf{0.493} \\
        \bottomrule
    \end{tabular}
    \begin{tablenotes}
        \footnotesize
        \item[*] Bold values indicate the best performance.
    \end{tablenotes}
    \label{tab:ratingpred}
    \end{threeparttable}
\end{table}

\subsection{Evaluation Metrics}
We evaluate rating prediction using RMSE and MAE, which measure the difference between the predicted and actual ratings. For explanation generation, we use BLEU-1, BLEU-4, ROUGE-2, and ROUGE-L (F1 scores) to assess similarity with ground truth text. Higher scores indicate better alignment.

\subsection{Baseline Methods}
Performances of each of the tasks are compared with the state-of-the-art models discussed below.

\subsubsection{Rating Prediction}

\textbf{PMF} \cite{mnih2007probabilistic} is a probabilistic matrix factorization model that learns the latent factors of users and items. Both \textbf{NARRE} \cite{chen2018neural} and \textbf{DAML} \cite{liu2019daml} leverage textual information to learn the user and item representations by modeling the user and item documents. \textbf{RGCL} \cite{shuai2022review} advances the user-item representation by exploiting the advantage of graph contrastive learning, demonstrating superior performance over several state-of-the-art models. NRT \cite{li2017neural}, CAML \cite{kang2023distillation}, PETER \cite{li2021personalized}, and SERMON \cite{liao2025aspect} focus on both rating prediction and explanation generation; therefore, we discuss them in the following sub-section to avoid repetition.

\subsubsection{Explanation Generation}
\textbf{NRT} \cite{li2017neural} uses deep learning to learn the user and item representation from IDs, leveraging them for rating prediction and tip generation. \textbf{CAML} \cite{kang2023distillation} uses historical reviews of users to represent their preferences and employs co-attention mechanisms to identify the most relevant reviews and concepts, integrating these concepts to generate explanation texts.  \textbf{ReXPlug} \cite{radford2019language} takes advantage of the large language model, e.g., GPT-2 to generate explanation text while predicting the ratings. Similarly, \textbf{PETER} \cite{li2021personalized} is a transformer-based explanation generation model that incorporates user and item IDs as tokens to the input sequence, demonstrating enhanced performance. Following that, \textbf{PEPLER} \cite{li2023personalized} builds on a pre-trained transformer model to produce explainable recommendations by using prompts that combine user and item ID vectors. To bridge the gap between these prompts and the pre-trained model, the method introduces sequential tuning and recommendation-based regularization techniques. \textbf{SERMON} \cite{liao2025aspect} is a powerful explanation generation model that considers aspects as context when generating explanations.

\begin{table*}[ht]
    \centering
    \footnotesize 
    \renewcommand{\arraystretch}{1.25} 
    \setlength{\tabcolsep}{5pt} 
    \caption{Performance comparison on explanation generation tasks using BLEU and ROUGE scores}
    \begin{tabular}{l|cccc|cccc|cccc}
        \toprule[1pt]
        \addlinespace[1pt]
        \multirow{2}{*}{\textbf{Models}} 
        & \multicolumn{4}{c|}{\textbf{Yelp}} 
        & \multicolumn{4}{c|}{\textbf{TripAdvisor}} 
        & \multicolumn{4}{c}{\textbf{Amazon}} \\
        \addlinespace[2pt]
        \cline{2-13}
        \addlinespace[2pt]
        & BLEU-1 & BLEU-4 & R-2 & R-L 
        & BLEU-1 & BLEU-4 & R-2 & R-L 
        & BLEU-1 & BLEU-4 & R-2 & R-L \\
        \addlinespace[2pt]
        \midrule[0.8pt]
        NRT       & 10.5 & 0.67 & 1.35 & 12.25  & 15.78 & 0.85 & 1.90 & 12.25  & 13.37 & 1.44 & 1.97 & 10.77 \\
        CAML      & 9.91 & 0.56 & 1.25 & 12.39  & 14.43 & 0.86 & 1.92 & 12.39  & 11.19 & 1.12 & 1.24 & 8.11 \\
        ReXPlug   & 8.59 & 0.57 & 1.11 & 9.97   & 12.64 & 0.71 & 1.61 & 9.97   & 10.80 & 1.29 & 1.22 & 8.73 \\
        PETER     & 10.29 & 0.69 & 1.43 & 12.61  & 15.33 & 0.89 & 1.94 & 12.61  & 13.78 & 1.68 & 1.97 & 11.07 \\
        PEPLER    & 10.42 & 0.73 & 1.46 & 11.64  & 15.62 & 1.09 & 2.22 & 11.64  & 13.78 & 1.05 & 2.22 & \textbf{13.19} \\
        SERMON    & 10.66 & 0.76 & 1.60 & 10.87  & \textbf{16.69} & 1.18 & 2.49 & \textbf{13.63} & 14.13 & \textbf{1.92} & 2.74 & 12.29 \\
        CLARER    & \textbf{12.60} & \textbf{7.16} & \textbf{7.90} & \textbf{14.31} 
                  & 11.52 & \textbf{4.79} & \textbf{4.53} & 11.70 
                  & \textbf{14.46} & 1.12 & \textbf{5.12} & 12.30 \\
        \addlinespace[2pt]
        \bottomrule[1pt]
    \end{tabular}
    \label{tab:performance_comparison_flipped}
\end{table*}

\subsection{Experimental details and Reproducibility}



We choose the \texttt{all-MiniLM-L6-v2} Sentence Transformer due to its lightweight architecture and competitive performance, and its 384-dimensional embeddings offers low inference latency. The multitask weights (1.0 for explanation and contrastive tasks, 0.5 for rating prediction) are empirically set to prioritize tasks with more complex objectives while maintaining contribution from all learning signals. The feedforward layer configuration of $[256 \rightarrow 128 \rightarrow 64]$, inspired by \cite{hasan2022multi}, provides a compact yet expressive projection space for joint representations. We adopt 6 Transformer layers and 8 attention heads to ensure sufficient model capacity without incurring excessive computational cost. The hidden dimension of 1024 is selected based on standard practice for mini versions of the Transformer model, ensuring compatibility with pretrained components. 

Adam optimizer with a learning rate of 0.001 and L2 regularization coefficient of 0.1 is chosen based on its robust convergence properties in similar multi-objective learning. The batch size is selected as 64 to balance memory usage and convergence stability, and early stopping (patience of 3 epochs) is used to prevent overfitting and reduce unnecessary computation.

\section{Results}
\subsection{Rating Prediction Accuracy}

Table \ref{tab:ratingpred} presents the performance of our proposed CLARER model in rating prediction tasks, compared to models that focus exclusively on rating prediction, as well as those that jointly address rating prediction and explanation generation problems. Our model consistently outperforms both types of models in terms of RMSE and MAE across all three datasets, achieving the best performance in all evaluations. The results show that our model achieves the highest performance improvement over the latest state-of-the-art model, SERMON \cite{liao2025aspect}, in the largest data set (Yelp), reducing RMSE by 29\% and MAE by 35\%, demonstrating substantial gains in the accuracy of rating prediction. The smallest improvement is observed in the TripAdvisor dataset, with 14\% lower RMSE and 16\% lower MAE compared to SERMON, while the Amazon dataset shows 19\% and 18\% reductions in RMSE and MAE, respectively. These results highlight our model's effectiveness across different domains, with particularly strong performance on larger datasets. 

The results also indicate that the prediction errors from review-based models (NARRE, DAML, RGCL) are much lower than rating-based models (PMF), demonstrating the effectiveness of review features. Furthermore, the recent explainable recommendation model beats both the rating- and review-based baselines in the rating prediction tasks, indicating the effectiveness of the advanced methodologies they adopted.

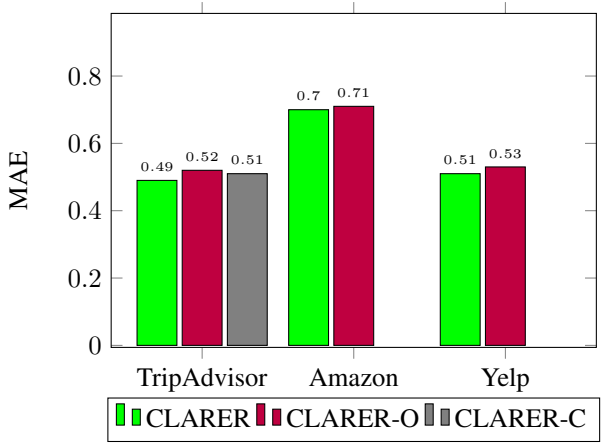
\begin{figure}
    \centering
    \begin{tikzpicture}
        \begin{axis}[
            ybar,
            ymin=0.18, ymax=0.8,
            bar width=15pt,
            enlargelimits=0.3,
            symbolic x coords={TripAdvisor, Amazon, Yelp},
            xtick=data,
            ylabel={MAE},
            legend style={at={(0.5,-0.15)}, anchor=north, legend columns=-1},
            nodes near coords,
            every node near coord/.append style={font=\tiny},
            width=8.0cm,
            height=6cm
        ]
            \addplot[fill=green] coordinates {(TripAdvisor, 0.49) (Amazon, 0.70) (Yelp, 0.51)};
            \addplot[fill=purple] coordinates {(TripAdvisor, 0.52) (Amazon, 0.71) (Yelp, 0.53)};
            \addplot[fill=gray] coordinates {(TripAdvisor, 0.51)};
            
            \legend{CLARER, CLARER-O, CLARER-C}
        \end{axis}
    \end{tikzpicture}
    \caption{MAE comparison across CLARER variants}
    \label{fig:ablation1}
\end{figure}

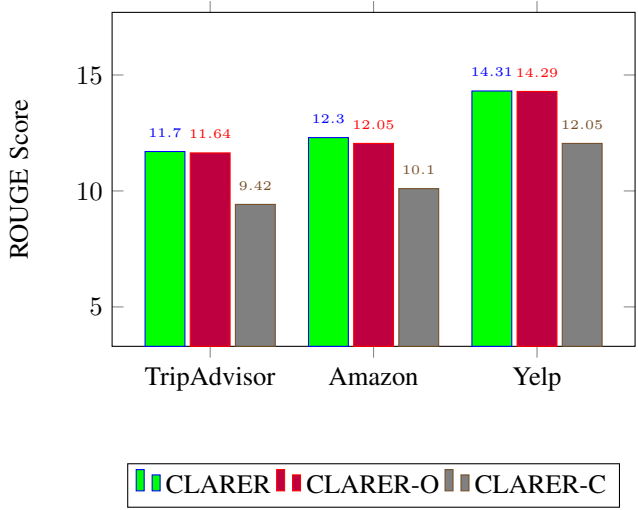
\begin{figure}[h]
    \centering
    \begin{tikzpicture}
        \begin{axis}[
            ybar,
            bar width=15pt,
            enlargelimits=0.3,
            ymin=6, ymax=15,
            symbolic x coords={TripAdvisor, Amazon, Yelp},
            xtick=data,
            ylabel={ROUGE Score},
            legend style={at={(0.5,-0.35)}, anchor=north, legend columns=-1},
            nodes near coords,
            nodes near coords align={vertical},
            every node near coord/.append style={font=\tiny, yshift=2pt},
            width=8.5cm,
            height=6.0cm,
        ]
            \addplot+[fill=green]  coordinates {(TripAdvisor, 11.70) (Amazon, 12.30) (Yelp, 14.31)};
            \addplot+[fill=purple] coordinates {(TripAdvisor, 11.64) (Amazon, 12.05) (Yelp, 14.29)};
            \addplot+[fill=gray]   coordinates {(TripAdvisor, 9.42) (Amazon, 10.10) (Yelp, 12.05)};
            \legend{CLARER, CLARER-O, CLARER-C}
        \end{axis}
    \end{tikzpicture}
    \caption{ROUGE score comparison across CLARER variants}
    \label{fig:ablation2}
\end{figure}

\begin{table*}[ht]
    \centering
    \footnotesize
    \renewcommand{\arraystretch}{1.25} 
    \setlength{\tabcolsep}{6pt} 
    \caption{Ablation studies of different feature-based performance on explanation generation}
    \resizebox{\textwidth}{!}{%
    \begin{tabular}{l|cccc|cccc|cccc}
        \toprule[1pt]
        \addlinespace[2pt]
        \multirow{2}{*}{\textbf{Models}} 
        & \multicolumn{4}{c|}{\textbf{Yelp}} 
        & \multicolumn{4}{c|}{\textbf{TripAdvisor}} 
        & \multicolumn{4}{c}{\textbf{Amazon}} \\
        \addlinespace[2pt]
        \cline{2-13}
        \addlinespace[2pt]
        & BLEU-1 & BLEU-4 & R-2 & R-L 
        & BLEU-1 & BLEU-4 & R-2 & R-L 
        & BLEU-1 & BLEU-4 & R-2 & R-L \\
        \addlinespace[2pt]
        \midrule[0.8pt]
        CLARER-O 
        & 12.94 & 7.12 & 7.68 & 14.29
        & 11.33 & 4.69 & 4.52 & 11.64
        & 11.32 & 4.05 & 4.56 & 12.03 \\
        \addlinespace[1pt]
        CLARER-C 
        & -- & -- & -- & --
        & 9.77 & 1.83 & 1.35 & 9.42
        & -- & -- & -- & -- \\
        \addlinespace[1pt]
        CLARER
        & 12.60 & \textbf{7.16} & \textbf{7.90} & \textbf{14.31}
        & \textbf{11.52} & \textbf{4.79} & \textbf{4.53} & \textbf{11.70}
        & \textbf{14.46} & \textbf{4.47} & \textbf{5.12} & \textbf{12.30} \\
        \addlinespace[2pt]
        \bottomrule[1pt]
    \end{tabular}
    }
    \label{tab:clarer_performance_selected}
\end{table*}

\subsection{Explanation Generation}
The performance of CLARER in explanation generation tasks is presented in Table \ref{tab:performance_comparison_flipped}. Our CLARER model demonstrates strong and consistent performance across three datasets when evaluated using BLEU1, BLEU4, ROUGE-2 (R2), and ROUGE-L (RL) metrics. In general, CLARER outperforms the selected baselines, achieving the highest scores in most metrics and datasets, especially excelling in the ROUGE-based metric (e.g., R2), which emphasizes the fluency and structural similarity of generated explanations. For example, CLARER achieves the best R2 scores in all three datasets, the best RL score on Yelp, and the second-best RL score on Amazon, indicating its strength in generating contextually rich and coherent explanations. CLARER outperforms all baselines in all evaluation metrics in the Yelp datasets. It also achieves the highest BLEU1 scores in the Amazon datasets and BLEU4 in the TripAdvisor datasets.

\subsection{Ablation Studies}

We perform ablation studies on both rating prediction and explanation generation tasks by evaluating two variants of our proposed framework. The first variant, CLARER-O, serves as the base model, which utilizes overall rating-based and review-based features without considering criteria or aspects for user and item representations. The second variant, CLARER-C, extends this by incorporating criteria-level rating features alongside the overall ratings to improve the quality of the representation. The original CLARER model is the most comprehensive, which further integrates aspect-based features with existing rating-based representations, aiming to capture more fine-grained user preferences and item characteristics for improved performance. To implement CLARER-O, we keep the MLP-based rating prediction component and the review feature component, removing the upper part (anything above the review feature) shown in Figure \ref{fig: architecture}. To implement CLARER-C, we combine criteria ratings with overall ratings for rating prediction (no aspect representation with contrastive learning, as shown in Figure \ref{fig: architecture}). Finally, for CLARER, we replace the criteria ratings with aspects as input to the contrastive learning component. 

For the Amazon and Yelp datasets, criteria-level ratings are not available; therefore, we evaluated only the overall rating-based and aspect-based feature representations for these datasets. As shown in Figure \ref{fig:ablation1}, across the three datasets, incorporating aspect-based representations consistently leads to lower MAE compared to the other two variants, highlighting the effectiveness of our proposed aspect representation learning method. Specifically, CLARER achieves the lowest MAE values of 0.49, 0.70, and 0.51 on the TripAdvisor, Amazon, and Yelp datasets, respectively. In contrast, using only overall rating features results in the highest prediction errors. When criteria-level ratings are available (on TripAdvisor), integrating them also leads to a lower MAE, confirming their usefulness, although they remain slightly less effective than aspect-based features. Similarly, as shown in Figure \ref{fig:ablation2} for explanation generation, ROUGE scores (RL) indicate that our proposed CLARER outperforms the other two variants. The highest ROUGE score is achieved on the Yelp dataset, while the lowest is reported on the TripAdvisor dataset from CLARER-O.

Table \ref{tab:clarer_performance_selected} presents more detailed comparative results of the ablation study on explanation generation in three data sets. Performance is measured using BLEU scores, which evaluate n-gram overlaps, as well as ROUGE-2 and ROUGE-L scores. We can see that the original CLARER model outperforms the two variants in most cases across three datasets (only slightly worse on BLEU1 for Yelp), suggesting that the incorporation of aspect-level information plays a crucial role in enhancing the fluency, relevance, and overall quality of the explanations.

\subsection{Discussion}
Using contrastive learning for aspect-based representation, CLARER effectively captures nuanced user and item features, resulting in more accurate and personalized predictions. The contrastive objective encourages the model to learn discriminative and aspect-aware review embeddings, which, in turn, facilitates a more precise alignment between user preferences and item features. This aspect-centric alignment is critical for improving recommendation quality in multi-criteria settings.

Despite its strong performance in rating prediction, CLARER does not consistently outperform all baselines in the explanation generation task across the three datasets.
In the Yelp dataset, CLARER achieves the best performance across all evaluation metrics, demonstrating the strength of its aspect-aware representations. These results suggest that the effectiveness of aspect-based modeling is domain-dependent and that careful selection of aspect terms is critical for accurate user and item representation learning. 

BLEU-1 scores are lower on the TripAdvisor dataset, likely because the reviews in this dataset are longer and more detailed, using a wider variety of words (e.g., talking about service, location, cleanliness, etc.). BLEU-4 scores are lower in the Amazon dataset, possibly because reviews in this dataset cover many different product types, making it harder for models to generate exact 4-word sequences that match the reference. ROUGE-L scores are moderate in the TripAdvisor dataset, possibly because the models find it challenging to follow the structured style of the reviews in this dataset. For Amazon, the big differences in review length and writing style between categories make it even harder for models to match the structure of the reference explanations, leading to lower ROUGE-L scores. Future work could explore alternative sets of aspects tailored to specific domains/datasets to enhance generalization. In addition, selecting a more advanced embedding model for aspect representation - potentially a larger transformer-based encoder - could further improve the quality of generated explanations.

\section{Conclusion}
In this paper, we introduce CLARER, a novel recommendation framework that combines user- and item-ID-based features with aspect-based review representation to improve both recommendation accuracy and explanation generation. Experimental results demonstrate that CLARER outperforms existing state-of-the-art recommendation models across multiple benchmark datasets, highlighting its ability to generate precise and interpretable recommendations. Furthermore, the model's contrastive learning framework significantly improves the robustness of representations, reducing noise, and enhancing generalization. 

Our work contributes to the growing body of research on explainable AI in recommendation systems by providing users with meaningful and interpretable justifications for recommendations, enhancing the trust and transparency of the system. A potential future direction could be the integration of multimodal data sources, such as images and user interaction logs, which may further refine the quality of the recommendation and enrich the interpretability. 

\section{Acknowledgment}
The authors sincerely appreciate the reviewers for their invaluable efforts and constructive feedback.

\begingroup
\scriptsize
\bibliographystyle{IEEEtran}
\bibliography{library}

@inproceedings{sarwar2001item,
  title={Item-based collaborative filtering recommendation algorithms},
  author={Sarwar, Badrul and Karypis, George and Konstan, Joseph and Riedl, John},
  booktitle={Proceedings of the 10th international conference on World Wide Web},
  pages={285--295},
  year={2001}
}

@inproceedings{he2017neural,
  title={Neural collaborative filtering},
  author={He, Xiangnan and Liao, Lizi and Zhang, Hanwang and Nie, Liqiang and Hu, Xia and Chua, Tat-Seng},
  booktitle={Proceedings of the 26th international conference on world wide web},
  pages={173--182},
  year={2017}
}

@article{li2023personalized,
  title={Personalized prompt learning for explainable recommendation},
  author={Li, Lei and Zhang, Yongfeng and Chen, Li},
  journal={ACM Transactions on Information Systems},
  volume={41},
  number={4},
  pages={1--26},
  year={2023},
  publisher={ACM New York, NY, USA}
}

@article{li2021personalized,
  title={Personalized transformer for explainable recommendation},
  author={Li, Lei and Zhang, Yongfeng and Chen, Li},
  journal={arXiv preprint arXiv:2105.11601},
  year={2021}
}

@inproceedings{li2020generate,
  title={Generate neural template explanations for recommendation},
  author={Li, Lei and Zhang, Yongfeng and Chen, Li},
  booktitle={Proceedings of the 29th ACM International Conf. on Information \& Knowledge Management},
  pages={755--764},
  year={2020}
}

@inproceedings{dong2017learning,
  title={Learning to generate product reviews from attributes},
  author={Dong, Li and Huang, Shaohan and Wei, Furu and Lapata, Maria and Zhou, Ming and Xu, Ke},
  booktitle={15th EACL 2017 Software Demonstrations},
  pages={623--632},
  year={2017},
  organization={ACL}
}

@incollection{tintarev2015explaining,
  title={Explaining recommendations: Design and evaluation},
  author={Tintarev, Nava and Masthoff, Judith},
  booktitle={Recommender systems handbook},
  pages={353--382},
  year={2015},
  publisher={Springer}
}

@article{zhang2020explainable,
  title={Explainable recommendation: A survey and new perspectives},
  author={Zhang, Yongfeng and Chen, Xu and others},
  journal={Foundations and Trends{\textregistered} in Information Retrieval},
  volume={14},
  number={1},
  pages={1--101},
  year={2020},
  publisher={Now Publishers, Inc.}
}

@article{chen2021generate,
  title={Generate natural language explanations for recommendation},
  author={Chen, Hanxiong and Chen, Xu and Shi, Shaoyun and Zhang, Yongfeng},
  journal={arXiv preprint arXiv:2101.03392},
  year={2021}
}

@inproceedings{chen2019personalized,
  title={Personalized fashion recommendation with visual explanations based on multimodal attention network: Towards visually explainable recommendation},
  author={Chen, Xu and Chen, Hanxiong and Xu, Hongteng and Zhang, Yongfeng and Cao, Yixin and Qin, Zheng and Zha, Hongyuan},
  booktitle={Proceedings of the 42nd International ACM SIGIR Conference on Research and Development in Information Retrieval},
  pages={765--774},
  year={2019}
}

@article{li2021caesar,
  title={Caesar: context-aware explanation based on supervised attention for service recommendations},
  author={Li, Lei and Chen, Li and Dong, Ruihai},
  journal={Journal of Intelligent Info. Sys.},
  volume={57},
  number={1},
  pages={147--170},
  year={2021},
  publisher={Springer}
}

@inproceedings{zhang2014explicit,
  title={Explicit factor models for explainable recommendation based on phrase-level sentiment analysis},
  author={Zhang, Yongfeng and Lai, Guokun and Zhang, Min and Zhang, Yi and Liu, Yiqun and Ma, Shaoping},
  booktitle={Proceedings of the 37th international ACM SIGIR conference on Research \& development in IR},
  pages={83--92},
  year={2014}
}

@article{zhang2019deep,
  title={Deep learning based recommender system: A survey and new perspectives},
  author={Zhang, Shuai and Yao, Lina and Sun, Aixin and Tay, Yi},
  journal={ACM computing surveys (CSUR)},
  volume={52},
  number={1},
  pages={1--38},
  year={2019},
  publisher={ACM New York, NY, USA}
}

@inproceedings{zheng2017joint,
  title={Joint deep modeling of users and items using reviews for recommendation},
  author={Zheng, Lei and Noroozi, Vahid and Yu, Philip S},
  booktitle={Proceedings of the tenth ACM Int. Conf. on Web Search and Data Mining},
  pages={425--434},
  year={2017}
}

@article{xu2021multi,
  title={Multi-modal discrete collaborative filtering for efficient cold-start recommendation},
  author={Xu, Yang and Zhu, Lei and Cheng, Zhiyong and Li, Jingjing and Zhang, Zheng and Zhang, Huaxiang},
  journal={IEEE Transactions on Knowledge and Data Engineering},
  volume={35},
  number={1},
  pages={741--755},
  year={2021},
  publisher={IEEE}
}

@inproceedings{li2017neural,
  title={Neural rating regression with abstractive tips generation for recommendation},
  author={Li, Piji and Wang, Zihao and Ren, Zhaochun and Bing, Lidong and Lam, Wai},
  booktitle={Proceedings of the 40th International ACM SIGIR conference on Research and Development in Information Retrieval},
  pages={345--354},
  year={2017}
}

@inproceedings{le2020synthesizing,
  title={Synthesizing aspect-driven recommendation explanations from reviews},
  author={Le, Trung-Hoang and Lauw, Hady W},
  year={2020},
  organization={IJCAI}
}

@article{liao2025aspect,
  title={Aspect-enhanced explainable recommendation with multi-modal contrastive learning},
  author={Liao, Hao and Wang, Shuo and Cheng, Hao and Zhang, Wei and Zhang, Jiwei and Zhou, Mingyang and Lu, Kezhong and Mao, Rui and Xie, Xing},
  journal={ACM Transactions on Intelligent Systems and Technology},
  volume={16},
  number={1},
  pages={1--24},
  year={2025},
  publisher={ACM New York, NY}
}

@inproceedings{wang2018explainable,
  title={Explainable recommendation via multi-task learning in opinionated text data},
  author={Wang, Nan and Wang, Hongning and Jia, Yiling and Yin, Yue},
  booktitle={The 41st international ACM SIGIR conference on research \& development in information retrieval},
  pages={165--174},
  year={2018}
}

@inproceedings{ma2019jointly,
  title={Jointly learning explainable rules for recommendation with knowledge graph},
  author={Ma, Weizhi and Zhang, Min and Cao, Yue and Jin, Woojeong and Wang, Chenyang and Liu, Yiqun and Ma, Shaoping and Ren, Xiang},
  booktitle={The world wide web conference},
  pages={1210--1221},
  year={2019}
}

@article{natarajan2020resolving,
  title={Resolving data sparsity and cold start problem in collaborative filtering recommender system using linked open data},
  author={Natarajan, Senthilselvan and Vairavasundaram, Subramaniyaswamy and Natarajan, Sivaramakrishnan and Gandomi, Amir H},
  journal={Expert Systems with Applications},
  volume={149},
  pages={113248},
  year={2020},
  publisher={Elsevier}
}

@inproceedings{seo2017interpretable,
  title={Interpretable convolutional neural networks with dual local and global attention for review rating prediction},
  author={Seo, Sungyong and Huang, Jing and Yang, Hao and Liu, Yan},
  booktitle={Proceedings of the eleventh ACM conference on recommender systems},
  pages={297--305},
  year={2017}
}

@article{tang2024gavamoe,
  title={GaVaMoE: Gaussian-Variational Gated Mixture of Experts for Explainable Recommendation},
  author={Tang, Fei and Shen, Yongliang and Zhang, Hang and Tan, Zeqi and Zhang, Wenqi and Hou, Guiyang and Song, Kaitao and Lu, Weiming and Zhuang, Yueting},
  journal={arXiv preprint arXiv:2410.11841},
  year={2024}
}

@inproceedings{chen2018neural,
  title={Neural attentional rating regression with review-level explanations},
  author={Chen, Chong and Zhang, Min and Liu, Yiqun and Ma, Shaoping},
  booktitle={Proceedings of the 2018 world wide web conference},
  pages={1583--1592},
  year={2018}
}

@inproceedings{liu2019daml,
  title={Daml: Dual attention mutual learning between ratings and reviews for item recommendation},
  author={Liu, Donghua and Li, Jing and Du, Bo and Chang, Jun and Gao, Rong},
  booktitle={Proceedings of the 25th ACM SIGKDD international conference on knowledge discovery \& data mining},
  pages={344--352},
  year={2019}
}

@inproceedings{dong2020asymmetrical,
  title={Asymmetrical hierarchical networks with attentive interactions for interpretable review-based recommendation},
  author={Dong, Xin and Ni, Jingchao and Cheng, Wei and Chen, Zhengzhang and Zong, Bo and Song, Dongjin and Liu, Yanchi and Chen, Haifeng and De Melo, Gerard},
  booktitle={Proceedings of the AAAI conference on artificial intelligence},
  volume={34},
  number={05},
  pages={7667--7674},
  year={2020}
}

@inproceedings{xia2019leveraging,
  title={Leveraging ratings and reviews with gating mechanism for recommendation},
  author={Xia, Haifeng and Wang, Zengmao and Du, Bo and Zhang, Lefei and Chen, Shuai and Chun, Gang},
  booktitle={Proceedings of the 28th ACM international conference on information and knowledge management},
  pages={1573--1582},
  year={2019}
}

@inproceedings{peng2020mutual,
  title={Mutual self attention recommendation with gated fusion between ratings and reviews},
  author={Peng, Qiyao and Liu, Hongtao and Yu, Yang and Xu, Hongyan and Dai, Weidi and Jiao, Pengfei},
  booktitle={Database Systems for Advanced Applications: 25th International Conference, DASFAA 2020, Jeju, South Korea, September 24--27, 2020, Proceedings, Part III 25},
  pages={540--556},
  year={2020},
  organization={Springer}
}

@inproceedings{lu2018coevolutionary,
  title={Coevolutionary recommendation model: Mutual learning between ratings and reviews},
  author={Lu, Yichao and Dong, Ruihai and Smyth, Barry},
  booktitle={Proceedings of the 2018 World Wide Web Conference},
  pages={773--782},
  year={2018}
}

@inproceedings{bao2014topicmf,
  title={Topicmf: Simultaneously exploiting ratings and reviews for recommendation},
  author={Bao, Yang and Fang, Hui and Zhang, Jie},
  booktitle={Proceedings of the AAAI conference on artificial intelligence},
  volume={28},
  number={1},
  year={2014}
}

@article{lee1999learning,
  title={Learning the parts of objects by non-negative matrix factorization},
  author={Lee, Daniel D and Seung, H Sebastian},
  journal={nature},
  volume={401},
  number={6755},
  pages={788--791},
  year={1999},
  publisher={Nature Publishing Group UK London}
}

@inproceedings{mcauley2013hidden,
  title={Hidden factors and hidden topics: understanding rating dimensions with review text},
  author={McAuley, Julian and Leskovec, Jure},
  booktitle={Proceedings of the 7th ACM conference on Recommender systems},
  pages={165--172},
  year={2013}
}

@inproceedings{diao2014jointly,
  title={Jointly modeling aspects, ratings and sentiments for movie recommendation (JMARS)},
  author={Diao, Qiming and Qiu, Minghui and Wu, Chao-Yuan and Smola, Alexander J and Jiang, Jing and Wang, Chong},
  booktitle={Proceedings of the 20th ACM SIGKDD international conference on Knowledge discovery and data mining},
  pages={193--202},
  year={2014}
}

@inproceedings{cheng2018aspect,
  title={Aspect-aware latent factor model: Rating prediction with ratings and reviews},
  author={Cheng, Zhiyong and Ding, Ying and Zhu, Lei and Kankanhalli, Mohan},
  booktitle={Proceedings of the 2018 world wide web conference},
  pages={639--648},
  year={2018}
}

@article{reimers2019sentence,
  title={Sentence-bert: Sentence embeddings using siamese bert-networks},
  author={Reimers, Nils and Gurevych, Iryna},
  journal={arXiv preprint arXiv:1908.10084},
  year={2019}
}

@article{vaswani2017attention,
  title={Attention is all you need},
  author={Vaswani, Ashish and Shazeer, Noam and Parmar, Niki and Uszkoreit, Jakob and Jones, Llion and Gomez, Aidan N and Kaiser, {\L}ukasz and Polosukhin, Illia},
  journal={Advances in neural information processing systems},
  volume={30},
  year={2017}
}

@inproceedings{shuai2022review,
  title={A review-aware graph contrastive learning framework for recommendation},
  author={Shuai, Jie and Zhang, Kun and Wu, Le and Sun, Peijie and Hong, Richang and Wang, Meng and Li, Yong},
  booktitle={Proceedings of the 45th international ACM SIGIR conference on research and development in IR},
  pages={1283--1293},
  year={2022}
}

@article{mnih2007probabilistic,
  title={Probabilistic matrix factorization},
  author={Mnih, Andriy and Salakhutdinov, Russ R},
  journal={Advances in neural information processing systems},
  volume={20},
  year={2007}
}

@inproceedings{kang2023distillation,
  title={Distillation from heterogeneous models for top-k recommendation},
  author={Kang, SeongKu and Kweon, Wonbin and Lee, Dongha and Lian, Jianxun and Xie, Xing and Yu, Hwanjo},
  booktitle={Proceedings of the ACM Web Conference 2023},
  pages={801--811},
  year={2023}
}

@article{radford2019language,
  title={Language models are unsupervised multitask learners},
  author={Radford, Alec and Wu, Jeffrey and Child, Rewon and Luan, David and Amodei, Dario and Sutskever, Ilya and others},
  journal={OpenAI blog},
  volume={1},
  number={8},
  pages={9},
  year={2019}
}

@inproceedings{chen2017explaining,
  title={Explaining recommendations based on feature sentiments in product reviews},
  author={Chen, Li and Wang, Feng},
  booktitle={Proceedings of the 22nd international conference on intelligent user interfaces},
  pages={17--28},
  year={2017}
}

@inproceedings{hasan2022multi,
  title={Multi-criteria rating and review based recommendation model},
  author={Hasan, Emrul and Ding, Chen and Cuzzocrea, Alfredo},
  booktitle={2022 IEEE International Conference on Big Data (Big Data)},
  pages={5494--5503},
  year={2022},
  organization={IEEE}
}

@article{le2020contrastive,
  title={Contrastive representation learning: A framework and review},
  author={Le-Khac, Phuc H and Healy, Graham and Smeaton, Alan F},
  journal={Ieee Access},
  volume={8},
  pages={193907--193934},
  year={2020},
  publisher={IEEE}
}
\endgroup

\end{document}